\documentclass[prd,nofootinbib,reprint,twocolumn,showpacs,superscriptaddress,floatfix]{revtex4-1}
\pdfoutput=1 
\usepackage{bm, color} 
\usepackage{amssymb,amsfonts,slashed,amsthm,amsmath, graphicx, soul, empheq, cancel, braket}
\usepackage{hyperref}
\usepackage[caption=false]{subfig}

\begin{document}

\newcommand{\vev}[1]{ \left\langle {#1} \right\rangle }
\newcommand{\eV}{ \ {\rm eV} }
\newcommand{\KeV}{ \ {\rm keV} }
\newcommand{\MeV}{\  {\rm MeV} }
\newcommand{\GeV}{\  {\rm GeV} }
\newcommand{\TeV}{\  {\rm TeV} }
\newcommand{\1}{\mbox{1}\hspace{-0.25em}\mbox{l}}
\newcommand{\Red}[1]{{\color{red} {#1}}}

\newcommand{\lmk}{\left(}  
\newcommand{\rmk}{\right)}
\newcommand{\lkk}{\left[}  
\newcommand{\rkk}{\right]}
\newcommand{\lhk}{\left \{ }  
\newcommand{\rhk}{\right \} }
\newcommand{\del}{\partial}  
\newcommand{\la}{\left\langle} 
\newcommand{\ra}{\right\rangle}
\newcommand{\half}{\frac{1}{2}}

\newcommand{\bea}{\begin{array}}
\newcommand{\eea}{\end{array}}
\newcommand{\beq}{\begin{eqnarray}}
\newcommand{\eeq}{\end{eqnarray}}
\newcommand{\eq}[1]{Eq.~(\ref{#1})}

\newcommand{\dd}{\mathrm{d}}
\newcommand{\Mpl}{M_{\rm Pl}}
\newcommand{\mg}{m_{3/2}}
\newcommand{\abs}[1]{\left\vert {#1} \right\vert}
\newcommand{\mphi}{m_{\phi}}
\newcommand{\Hz}{\ {\rm Hz}}
\newcommand{\for}{\quad \text{for }}
\newcommand{\Min}{\text{Min}}
\newcommand{\Max}{\text{Max}}
\newcommand{\Kahler}{K\"{a}hler }
\newcommand{\cphi}{\varphi}
\newcommand{\Tr}{\text{Tr}}
\newcommand{\diag}{{\rm diag}}

\newcommand{\SUf}{SU(3)_{\rm f}}
\newcommand{\Upq}{U(1)_{\rm PQ}}
\newcommand{\Zpq}{Z^{\rm PQ}_3}
\newcommand{\Cpq}{C_{\rm PQ}}
\newcommand{\ubar}{u^c}
\newcommand{\dbar}{d^c}
\newcommand{\ebar}{e^c}
\newcommand{\nubar}{\nu^c}
\newcommand{\Ndw}{N_{\rm DW}}
\newcommand{\Fpq}{F_{\rm PQ}}
\newcommand{\fpq}{v_{\rm PQ}}
\newcommand{\Br}{{\rm Br}}
\newcommand{\Lag}{\mathcal{L}}
\newcommand{\Lqcd}{\lambda_{\phi}_{\rm QCD}}

\newcommand{\ji}{j_{\rm inf}} 
\newcommand{\jb}{j_{B-L}} 
\newcommand{\M}{M} 
\newcommand{\im}{{\rm Im} }
\newcommand{\re}{{\rm Re} }

\def\lrf#1#2{ \left(\frac{#1}{#2}\right)}
\def\lrfp#1#2#3{ \left(\frac{#1}{#2} \right)^{#3}}
\def\lrp#1#2{\left( #1 \right)^{#2}}
\def\REF#1{Ref.~\cite{#1}}
\def\SEC#1{Sec.~\ref{#1}}
\def\FIG#1{Fig.~\ref{#1}}
\def\EQ#1{Eq.~(\ref{#1})}
\def\EQS#1{Eqs.~(\ref{#1})}
\def\TEV#1{10^{#1}{\rm\,TeV}}
\def\GEV#1{10^{#1}{\rm\,GeV}}
\def\MEV#1{10^{#1}{\rm\,MeV}}
\def\KEV#1{10^{#1}{\rm\,keV}}
\def\blue#1{\textcolor{blue}{#1}}
\def\red#1{\textcolor{blue}{#1}}

\title{Strong First-Order Phase Transition in a Pseudo-Nambu-Goldstone Boson Model}

\author{Yuichiro Nakai}
\affiliation{Tsung-Dao Lee Institute, Shanghai Jiao Tong University, \\
No.~1 Lisuo Road, Pudong New Area, Shanghai 201210, China}
\affiliation{School of Physics and Astronomy, Shanghai Jiao Tong University, \\
800 Dongchuan Road, Shanghai 200240, China}
\author{Masanori Tanaka}
\affiliation{Tsung-Dao Lee Institute, Shanghai Jiao Tong University, \\
No.~1 Lisuo Road, Pudong New Area, Shanghai 201210, China}
\affiliation{School of Physics and Astronomy, Shanghai Jiao Tong University, \\
800 Dongchuan Road, Shanghai 200240, China}
\author{Yaoduo Wang}
\affiliation{Tsung-Dao Lee Institute, Shanghai Jiao Tong University, \\
No.~1 Lisuo Road, Pudong New Area, Shanghai 201210, China}
\affiliation{School of Physics and Astronomy, Shanghai Jiao Tong University, \\
800 Dongchuan Road, Shanghai 200240, China}
\date{\today}

\begin{abstract}

We propose a novel, calculable model for realizing strong first-order phase transitions in the early Universe, in which the phase transition occurs along a pseudo-Nambu-Goldstone boson (pNGB) direction. The model involves two scalar fields whose interactions exhibit collective symmetry breaking, protecting the pNGB potential from quadratically divergent radiative corrections induced by these interactions. This structure allows strong first-order phase transitions to be realized without severe parameter tuning while maintaining perturbative control. We investigate the finite-temperature effective potential and bubble nucleation dynamics, identifying parameter regions where strong first-order phase transitions can successfully complete. The strength and characteristic time scale of the transitions are evaluated, and the resulting gravitational-wave (GW) spectra are computed. The predicted GW signals can fall within the sensitivity ranges of planned GW experiments.

\end{abstract}

\maketitle

\section{Introduction}

Cosmological first-order phase transitions can generate stochastic gravitational waves (GWs), providing a novel avenue for probing physics beyond the Standard Model (for reviews, see e.g. Refs.~\cite{Weir:2017wfa,Mazumdar:2018dfl,Caprini:2019egz,Hindmarsh:2020hop,Athron:2023xlk,Caprini:2024hue}). The amplitude of GWs is controlled by several properties of a first-order phase transition, most importantly the amount of energy available to source GWs and the characteristic time scale of the transition. These are conventionally characterized by the transition strength $\alpha$, defined as the released energy normalized by the radiation energy density, and the inverse duration $\beta$ in units of the Hubble expansion rate $H$. A large GW signal is favored by a strong transition, typically with $\alpha \gtrsim 1$, and a sufficiently slow phase transition with a small $\beta/H$. An important model-building question is therefore how to achieve such a strong and slow first-order phase transition naturally and within theoretical control.

Achieving a strong and slow first-order phase transition usually requires a particular structure of the scalar potential, whose stability against radiative corrections is generally not guaranteed. Classically scale-invariant theories
\cite{Witten:1980ez,Iso:2009ss,Iso:2009nw,Konstandin:2011dr,Jinno:2016knw,Hashino:2016rvx,Jaeckel:2016jlh,Iso:2017uuu,Marzo:2018nov,Hambye:2018qjv,Ellis:2019oqb,Ellis:2020nnr,Dasgupta:2022isg,Ahriche:2023jdq,Sagunski:2023ynd,Liu:2024fly,Cao:2025jwb,Rescigno:2025ong,Das:2026rwf}
provide a well-known framework in which dimensional transmutation generates a symmetry-breaking scale and is able to lead to a strong first-order phase transition. However, the absence of a tree-level scalar mass term, which is essential to the mechanism, is not protected against ultraviolet corrections by classical scale invariance. On the other hand, nearly conformal field theories and their warped extra-dimensional descriptions
\cite{Creminelli:2001th,vonHarling:2017yew,Baratella:2018pxi,Fujikura:2019oyi,Fujikura:2023lkn,Fujikura:2025iam,Agrawal:2025wvf}
can naturally realize strong first-order phase transitions, although quantitative predictions are often complicated by nonperturbative tunneling dynamics and gravitational effects. These considerations motivate the construction of a simple, weakly coupled model that realizes a strong and slow first-order phase transition with a radiatively stable scalar potential while allowing a perturbative analysis of its dynamics.

In this paper, we introduce a novel, calculable model in which a strong first-order phase transition takes place along the direction of a pseudo-Nambu-Goldstone boson (pNGB). In addition to the pNGB, the model contains a complex scalar field whose interactions exhibit collective symmetry breaking: each interaction separately preserves a continuous shift symmetry of the pNGB, while the symmetry is broken when both are present.\footnote{The mechanism of collective symmetry breaking has been known in the context of little Higgs models~\cite{Arkani-Hamed:2001nha,Arkani-Hamed:2002ikv,Schmaltz:2005ky}.} This structure protects the pNGB potential against quadratically divergent radiative corrections induced by these interactions. Another important feature of the model is the mass splitting between the two eigenstates of the complex scalar field. When both eigenstates are light compared to the temperature, their leading thermal contributions to the pNGB potential cancel. However, when the heavier state becomes Boltzmann suppressed, the lighter state generates a nonvanishing temperature-dependent potential~\cite{Espinosa:2004pn}. The interplay between this thermal effect, the zero-temperature radiative corrections, and a tree-level potential allows strong and slow first-order phase transitions to occur while maintaining radiative stability. The weakly coupled framework enables a perturbative analysis of the finite-temperature effective potential and bubble nucleation dynamics.

To examine the cosmological evolution, we determine the temperature dependence of the effective potential and numerically compute the thermal tunneling action. The nucleation rate is then utilized to determine whether the transition successfully completes. We map out the viable model parameter space and characterize the transitions in terms of their strength $\alpha$ and characteristic time scale $\beta^{-1}$. Using these results, we estimate a GW background generated during the transition.

The rest of the paper is organized as follows. Sec.~\ref{sec:model} introduces our model and discusses its finite-temperature effective potential and bubble nucleation. In Sec.~\ref{sec:numerical}, we present numerical results for the phase transitions, identify regions of parameter space where strong first-order phase transitions successfully complete, and evaluate the resulting GW spectra. In Sec.~\ref{sec:tuning}, we discuss the radiative stability of the scalar potential and examine the degree of parameter tuning required in our model. Finally, Sec.~\ref{sec:conclusion} is devoted to conclusions and discussions.

\section{The Model}
\label{sec:model}

Let us consider a model with a real singlet scalar field $\phi$ and a complex singlet scalar field $\sigma$, described by the following Lagrangian:
\begin{align}
\label{our_lagrangian}
\mathcal{L}
=& \frac{1}{2}(\partial_\mu\phi)^2 + |\partial_\mu \sigma|^2 -V_0(\phi, \sigma) \,. 
\end{align}
The tree-level scalar potential $V_{0}(\phi, \sigma)$ is given by
\begin{align}
\label{our_potential}
\begin{aligned}
V_{0}(\phi, \sigma) 
= ~ &  \mu_{\sigma}^2 |\sigma|^2 + \lambda_{\sigma} |\sigma|^4  + \left[  c(\phi) \sigma^2 + {\rm h.c.} \right] \\[1ex] 
&  + A \left[1  - \cos \left( \phi/f \right) \right] \,.\\
\end{aligned}
\end{align}
Here, $c(\phi)\equiv m_0^2+m_1^2e^{-i\phi/f}$, where $f$ is the decay constant of the pNGB $\phi$, which is associated with a spontaneously broken global $U(1)$ symmetry and transforms as $\phi\to\phi+\epsilon f$ under the corresponding continuous shift symmetry. The interactions involving $\sigma$ exhibit a collective symmetry-breaking structure: each of the two terms in $c(\phi)$ preserves a continuous shift symmetry, but with different charge assignments for $\sigma$. The term $m_0^2\sigma^2$ is invariant under the shift of $\phi$ alone, whereas $m_1^2e^{-i\phi/f}\sigma^2$ is invariant under the combined transformation $\phi\to\phi+\epsilon f$ and $\sigma\to e^{i\epsilon/2}\sigma$. Consequently, the shift symmetry is broken only when both couplings are present, so that radiative corrections to the $\phi$ potential arising from these interactions must involve both $m_0^2$ and $m_1^2$. This mechanism is similar to the little Higgs scenario~\cite{Arkani-Hamed:2001nha,Arkani-Hamed:2002ikv,Schmaltz:2005ky}. In contrast, the $A$ term explicitly breaks the continuous shift symmetry and introduces a potential for $\phi$ at the tree-level. The approximate shift symmetry thus controls the radiative corrections to the $\phi$ potential, protecting its structure against large quantum corrections.

To investigate the phase transition along the $\phi$ direction, we consider the finite-temperature effective
potential,
\begin{align}
\label{eq:VeffT}
\begin{aligned}
V(\phi, T)
=& V_{0}(\phi, \sigma = 0) + V_{\rm CW}(\phi) + V_{T}(\phi, T) \\
& - \frac{T^2}{24} \sum_{i = \sigma_\pm} \frac{\partial V_\sigma^{(0)}}{\partial m_i^2} \frac{\partial^2 m_i^2(\phi)}{\partial\phi^2} + V_{\rm ring}(\phi, T) \,,
\end{aligned}
\end{align}
where $V_{\rm CW}(\phi)$ and $V_{T}(\phi, T)$ denote the one-loop corrections at zero and finite temperatures~\cite{Coleman:1973jx, Dolan:1973qd}.
Each term is expressed as
\begin{align}
& V_{\rm CW}(\phi) = \sum_{i = \phi,\,\sigma_\pm} \frac{m_i^4(\phi)}{64\pi^2}\left[\ln\frac{m_i^2(\phi)}{\mu^2} - \frac{3}{2}\right] \,, \label{eq:VCW} \\
& V_{T}(\phi, T) = \sum_{i = \phi,\,\sigma_\pm} \frac{T^4}{2\pi^2}\,J_B\!\left(\frac{m_i^2(\phi)}{T^2}\right) \,, \label{eq:VT}
\end{align}
where the function $J_{B}(a^2)$ is defined by~\cite{Dolan:1973qd}
\begin{align}
J_B(a^2) = \int_{0}^{\infty} dx x^2 \ln \left[1 - e^{-\sqrt{x^2 + a^2}} \right] \,.
\end{align}
For the renormalization scale, we take $\mu = f$ in the subsequent analysis.
The masses $m_i^2$ and the last two terms in Eq.~\eqref{eq:VeffT} are specified below.

The mass eigenvalues for the $\sigma$ field at the point $\sigma = 0$ are given by
\begin{align}
\label{eq:sigma_mass}
m_{\sigma_\pm}^2(\phi) 
& = \mu_\sigma^2 \pm 2\,|c(\phi)| \nonumber \\[1ex]
& = \mu_\sigma^2 \pm 2\sqrt{m_0^4 + m_1^4 + 2 m_0^2 m_1^2 \cos(\phi/f)} \,,
\end{align}
where $\sigma_\pm$ represent the mass eigenstates for the $\sigma$ field.
Since $m_{\sigma_+}^2+m_{\sigma_-}^2=2\mu_\sigma^2$ is independent of $\phi$, the quadratically divergent contribution to the one-loop effective potential from the $\sigma$ scalar field does not depend on $\phi$, as expected from the collective symmetry-breaking structure. The same relation implies that the leading $\mathcal{O}(T^2)$ thermal corrections cancel when both of the mass eigenstates are light compared to the temperature. However, when the two eigenstates have sufficiently different masses, the thermal contribution from the heavier state is Boltzmann suppressed, leaving a nonvanishing $\phi$-dependent contribution from the lighter state~\cite{Espinosa:2004pn}. This mass hierarchy plays a role in generating the finite-temperature potential required for a strong first-order phase transition. In the following analysis, we focus on the case $m_0^2m_1^2<0$, where $\sigma_-$ is the lightest at $\phi=\pi f$, and take $m_1^2=-m_0^2$ for simplicity.

When evaluating $V_{\rm CW}$ and $V_T$, we include thermal corrections to the $\sigma$ masses arising from fluctuations of $\phi$. These corrections are particularly important when $\sigma_-$ is much lighter than the temperature around $\phi=\pi f$. Assuming that the heavier component satisfies $m_{\sigma_+}\gg T$, the self-energy corrections from the cubic and quartic interactions involving $\phi$ in $c(\phi)\sigma^2$ can be treated as local. Their combined effect can be incorporated by averaging the mass eigenvalues over thermal fluctuations $\delta\phi$, with $\langle\delta\phi^2\rangle=T^2/12$:
\begin{align}
\label{eq:sigma_mass_T}
M_{\sigma_\pm}^2(\phi, T)
& \equiv \braket{ m_{\sigma_\pm}^2(\phi + \delta\phi) } \nonumber \\
& = m_{\sigma_\pm}^2(\phi) + \frac{T^2}{24}\,\frac{\partial^2 m_{\sigma_\pm}^2(\phi)}{\partial\phi^2} \,. 
\end{align}
Taking the Gaussian average with the assumption $m_1^2 = -m_0^2$, we obtain $M_{\sigma_\pm}^2(\phi, T) = \mu_\sigma^2 \pm 2|c(\phi)| e^{-T^2/(96f^2)}$, so that $M_{\sigma_-}^2(\pi f, T) \simeq m^2_{\sigma_{-}}(\pi f) + |m_{0}^2| T^2 /(24f^2)$.

For the field dependent mass of $\phi$, we take into account the radiative correction from the $\sigma$ field because its effect is comparable to the tree-level contribution in our focused parameter space:
\begin{align}
\label{eq:mphi}
& m_\phi^2(\phi) 
=  \frac{\partial^2}{\partial\phi^2} \left[V_0(\phi, 0) + V_\sigma^{(0)}(\phi) \right] \,,
\\[1ex]
& V_\sigma^{(0)}(\phi) \equiv  \sum_{i = \sigma_{\pm}} \frac{m_{i}^4(\phi)}{64\pi^2}
\left[\ln\frac{m_{i}^2(\phi)}{\mu^2} - \frac{3}{2}\right] \,.
\end{align}
In Eq.~\eqref{eq:mphi}, the heavy $\sigma_+$ loop contribution to the $\phi$ propagator is incorporated by following the resummation prescription for Nambu-Goldstone bosons to avoid infrared divergences
\cite{Martin:2014bca,Elias-Miro:2014pca,Chiang:2018gsn}.
The fourth term in Eq.~\eqref{eq:VeffT} removes the two-loop contributions with a thermal $\phi$ line and vacuum $\sigma$ lines that are included both in the $\phi$ contribution to $V_T$ through $m_\phi^2$ and in $V_{\rm CW}$ through $M_{\sigma_\pm}^2$~\cite{Elias-Miro:2014pca}.

We also take into account the daisy resummation term $V_{\rm ring}(\phi, T)$ for the $\phi$ field based on the Arnold-Espinosa scheme~\cite{Arnold:1992rz}:
\begin{align}
V_{\rm ring}(\phi, T) 
= -\frac{T}{12\pi}\left[\big(M_\phi^2\big)^{3/2} - \big(m_\phi^2\big)^{3/2}\right] \,,
\end{align}
where the thermal mass of $\phi$ is given by
\begin{align}
& M_\phi^2(\phi, T) = m_\phi^2(\phi) + \Pi_\phi(\phi, T) \,, \\[1ex]
& \Pi_\phi(\phi, T)
= \frac{T^2}{24}\,\frac{\partial^4 V_0}{\partial\phi^4}
+ \frac{\partial^2}{\partial\phi^2}\Big[V_\sigma(\phi, T) - V_\sigma^{(0)}(\phi)\Big] \,.
\end{align}
Here $V_\sigma(\phi, T)$ denotes the contribution from the $\sigma$ field in $V_{\rm CW} + V_T$ with the masses in Eq.~\eqref{eq:sigma_mass_T}.
The first term of $\Pi_\phi$ is the conventional thermal mass correction~\cite{Carrington:1991hz}.
The second denotes the thermal curvature induced by the $\sigma$ loops, which cannot be replaced by its high-temperature form because the mass of $\sigma_{+}$ component is heavier than the typical temperature in our setup.
These energy scale hierarchies are important because the $\phi$ dependence disappears if $m_{\sigma_{\pm}} \ll T$ due to the $\phi$ dependence cancellation in the relation $m_{\sigma_+}^2 + m_{\sigma_-}^2$.
Where $m_\phi^2$ or $M_\phi^2$ is negative, we take the real parts of the corresponding terms in Eq.~\eqref{eq:VeffT}.

\begin{figure}[t]
    \centering
    \includegraphics[width=0.98\linewidth]{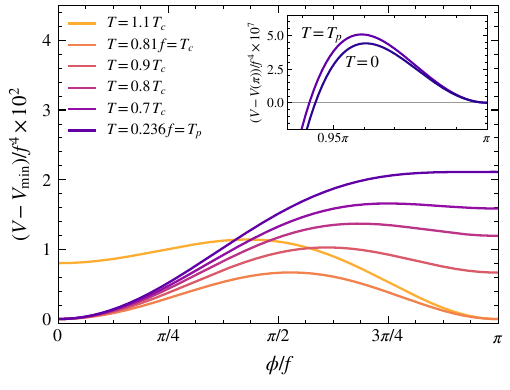}
    \caption{
    Temperature evolution of the effective potential~\eqref{eq:VeffT} for $f = 10^4\,\mathrm{GeV}$, $m_0^2 = -m_1^2 = 2f^2$, $\lambda_{\sigma} = 0.1$, $\mu_\sigma^2 = 2(|m_0^2| + |m_1^2|) + M_\pi^2$ with $M_\pi = f$, and $A/f^4 = -0.2038$.
    The upper-right small plot shows the potential structure around $\phi = \pi f$ at $T = T_p$ and $T = 0$.    
    }
    \label{fig:potential_evolution}
\end{figure}

Figure~\ref{fig:potential_evolution} describes the temperature evolution of the effective potential in Eq.~\eqref{eq:VeffT} for a benchmark point with $f = 10^4\,\mathrm{GeV}$, $m_0^2 = -m_1^2 = 2f^2$, $A/f^4 = -0.2038$, and $\mu_\sigma^2 = 2(|m_0^2| + |m_1^2|) + M_\pi^2$ with $M_\pi = f$. 
At high temperatures above the critical temperature $T_{c}$, the configuration $\phi = \pi f$ gives a global minimum potential energy. 
As the temperature decreases, the first-order phase transition occurs from the false vacuum $\phi = \pi f$ to the true vacuum $\phi = 0$ at the percolation temperature $T_{p}$. 
The precise definition of $T_{p}$ is given in the next section (see Eq.~\eqref{eq:Tp_Definition}). 
As shown in Fig.~\ref{fig:potential_evolution}, the potential structure around the point $\phi = \pi f$ at $T \leq T_{p}$ can be approximately flat due to the cancellation between the $A$ term and the loop correction from the $\sigma_{+}$ field. 
In this case, the small potential barrier around $\phi = \pi f$ is induced by the radiative correction from the $\sigma_{-}$ field whose thermal mass is given by $M_{\sigma_-}^2(\pi f, T) \simeq M_{\pi}^2 + |m_{0}^2|T^2/(24f^2)$. 
If $M_{\pi}^2 \ll |m_{0}^2| T^2/(24f^2)$, the potential structure around $\phi = \pi f$ is determined by the thermal mass of the $\sigma_{-}$ field. 
This situation is similar to that for models with classical scale invariance, which can lead to supercooled first-order phase transitions (see, e.g., Ref.~\cite{Iso:2017uuu}).

We calculate the bubble nucleation rate by using the finite-temperature effective potential \eqref{eq:VeffT}. At finite temperatures, thermal tunneling is governed by the three-dimensional Euclidean bounce action $S_3$, with the nucleation rate per unit volume approximately given by $\Gamma(T) \sim T^4 e^{-S_3(T)/T}$~\cite{Linde:1981zj}. Here, we numerically estimate $S_3(T)$ by using the public code \texttt{CosmoTransitions}~\cite{Wainwright:2011kj}.
The resulting nucleation rate is then used in the following section to analyze the phase-transition dynamics.

\section{Numerical results}
\label{sec:numerical}

In this section, we explore the cosmological phase transitions realized in our model through a numerical analysis of the finite-temperature effective potential in Eq.~\eqref{eq:VeffT}. We first map out the parameter space in which first-order phase transitions occur and successfully complete, paying attention to the possibility of supercooled transitions. We then evaluate the strength and duration of the phase transitions and discuss the resulting GW signals.

\subsection{Prediction for phase transition parameters}

We first discuss a theoretical constraint on first-order phase transitions, arising from the requirement that the transition must be completed by the present epoch.
To determine whether the phase transition completes, we employ the criterion~\cite{Turner:1992tz,Ellis:2019oqb,Athron:2022mmm},
\begin{align}
\label{eq:PT_completion}
{\rm CR}(T_{p}) \equiv \left. -T \frac{d I(T)}{dT} \right|_{T = T_p} > 3 \,,
\end{align}
where $I(T)$ is defined by~\cite{Turner:1992tz}
\begin{align}
I(T) 
= \frac{4\pi}{3} \int_{T}^{T_{c}} \frac{dT'}{T'^4} \frac{\Gamma(T')}{H(T')} \left( \int_{T}^{T'} d \tilde T \frac{v_{w}}{H( \tilde T)}\right)^3  \,,
\end{align}
where $T_{c}$ and $H(T)$ are the critical temperature and the Hubble parameter, respectively. 
The factor $v_{w}$ denotes the vacuum bubble wall velocity. 
In our analysis, we take $v_{w} = 1$ for simplicity. 
The percolation temperature $T_p$ is defined by the condition~\cite{Ellis:2019oqb},
\begin{align}
\label{eq:Tp_Definition}
I(T_p) = 0.34 \,.
\end{align}
The condition~\eqref{eq:PT_completion} implies that the volume of false vacuum domains should decrease after the percolation temperature. 
We consider the phase transition incomplete if the condition in Eq.~\eqref{eq:PT_completion} is not satisfied.

The two important physical quantities relevant to the dynamics of first-order phase transitions are the transition strength and the typical duration $\beta^{-1}$. 
The transition strength $\alpha$ is defined from the trace anomaly as~\cite{Grojean:2006bp, Athron:2023xlk}
\begin{align}
\label{eq:alpha_GW}
\alpha \equiv
\frac{1}{\rho_R(T_p)}
\left[
\Delta V(T_p)
-\frac{T_p}{4}
\left.
\frac{d\Delta V(T)}{dT}
\right|_{T=T_p}
\right] \,,
\end{align}
where $\Delta V$ denotes the potential difference between the false vacuum $\phi_{\rm false}$ and the true vacuum $\phi_{\rm true}$, and $\rho_{R}(T) = \pi^2 g_* T^4/30$ is the radiation energy density with $g_* = 109.75$, counting the Standard Model, $\phi$ and $\sigma$.
The inverse duration $\beta$ is defined by~\cite{Grojean:2006bp}
\begin{align}
\label{eq:beta_GW}
\beta \equiv \left. H(T)T\frac{d}{d T} \left( \frac{S_{3}}{T} \right) \right|_{T = T_{p}} \,.
\end{align}
For strongly supercooled phase transitions, the parameter $\beta$ can become small or even negative~\cite{Wang:2020jrd, Kanemura:2024pae}. 
In such a case, using $\beta$ to estimate the typical mean separation of vacuum bubbles is not appropriate. 
Instead, we define the typical mean separation length $R_{*}$ as~\cite{Ellis:2018mja, Wang:2020jrd, Kanemura:2024pae}
\begin{align}
R_{*} = n_{B}^{-1/3}(T_{p}) \,,
\end{align}
where $n_{B}(T)$ denotes the bubble number density at the temperature $T$, which is defined by
\begin{align}
n_{B}(t) = \int_{t_{c}}^{t} dt' \frac{a^3(t')}{a^3(t)} \Gamma(t') e^{-I(t')} \,, 
\end{align}
where $a(t)$ represents the scale factor. 
The time $t_{c}$ corresponds to the cosmological time at the critical temperature $T_c$. 
In estimating the produced GW spectrum, we use $\beta_{\rm eff}$ instead of $\beta$ in the fitting function: 
\begin{align}
\beta_{\rm eff} \equiv (8 \pi)^{1/3} \frac{v_{w}}{R_{*}} \,. 
\end{align}

\begin{figure}[t]
    \centering
    \includegraphics[width=0.98\linewidth]{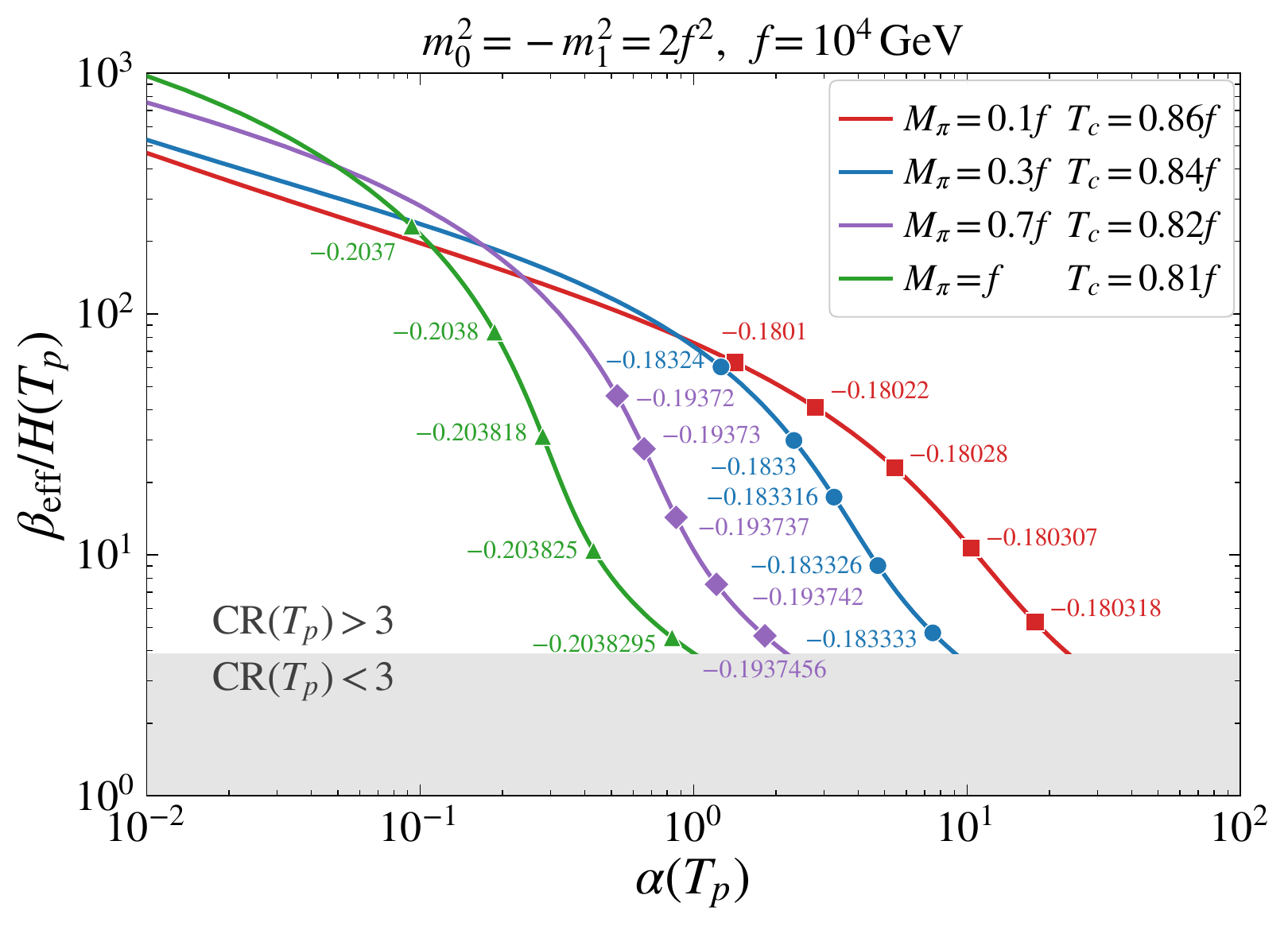}
    \caption{
    Parameter dependence of $\alpha$ and $\beta_{\rm eff}/H$ for different values of $A/f^4$, with $m_0^2= - m_1^2= 2 f^2$, $\lambda_{\sigma} = 0.1$ and $f = 10^{4}\,{\rm GeV}$. The labels indicate the values of $A/f^4$ at the corresponding points. Above the gray region, the phase transition completion condition in Eq.~\eqref{eq:PT_completion} is satisfied.  
    \label{fig:alpha_beta}}
\end{figure}

In Fig.~\ref{fig:alpha_beta}, the parameter value dependence of phase transition parameters is shown in the $(\alpha, \beta_{\rm eff}/H)$ plane with $m_{0}^2 = - m_{1}^2 = 2f^2$ and $f = 10^{4}\,{\rm GeV}$. 
The colored solid lines show the trajectory with the change in the $A$ value for each $M_{\pi}$ value. 
In the gray region, the completion condition for first-order phase transitions in Eq.~\eqref{eq:PT_completion} is not satisfied. 
Figure~\ref{fig:alpha_beta} indicates that a large transition strength $\alpha$ can be achieved as the value of $M_{\pi}$ decreases. 
This behavior is consistent with the discussion of Fig.~\ref{fig:potential_evolution}, where smaller values of $M_{\pi}$ favor supercooled first-order phase transitions. We note that the results shown in Fig.~\ref{fig:alpha_beta} are largely insensitive to the decay constant $f$, whose dependence enters only through the Hubble parameter.

\subsection{GW production}

A stochastic GW background can be generated by bubble dynamics during the first-order phase transition (see Ref.~\cite{Caprini:2024hue} and references therein). The resulting GW spectrum is primarily characterized by the parameters $\alpha$ and $\beta_{\rm eff}$. To evaluate the GW spectrum, we adopt the fitting formula summarized in Ref.~\cite{Caprini:2024hue}.

\begin{figure}[t]
    \centering
    \includegraphics[width=0.98\linewidth]{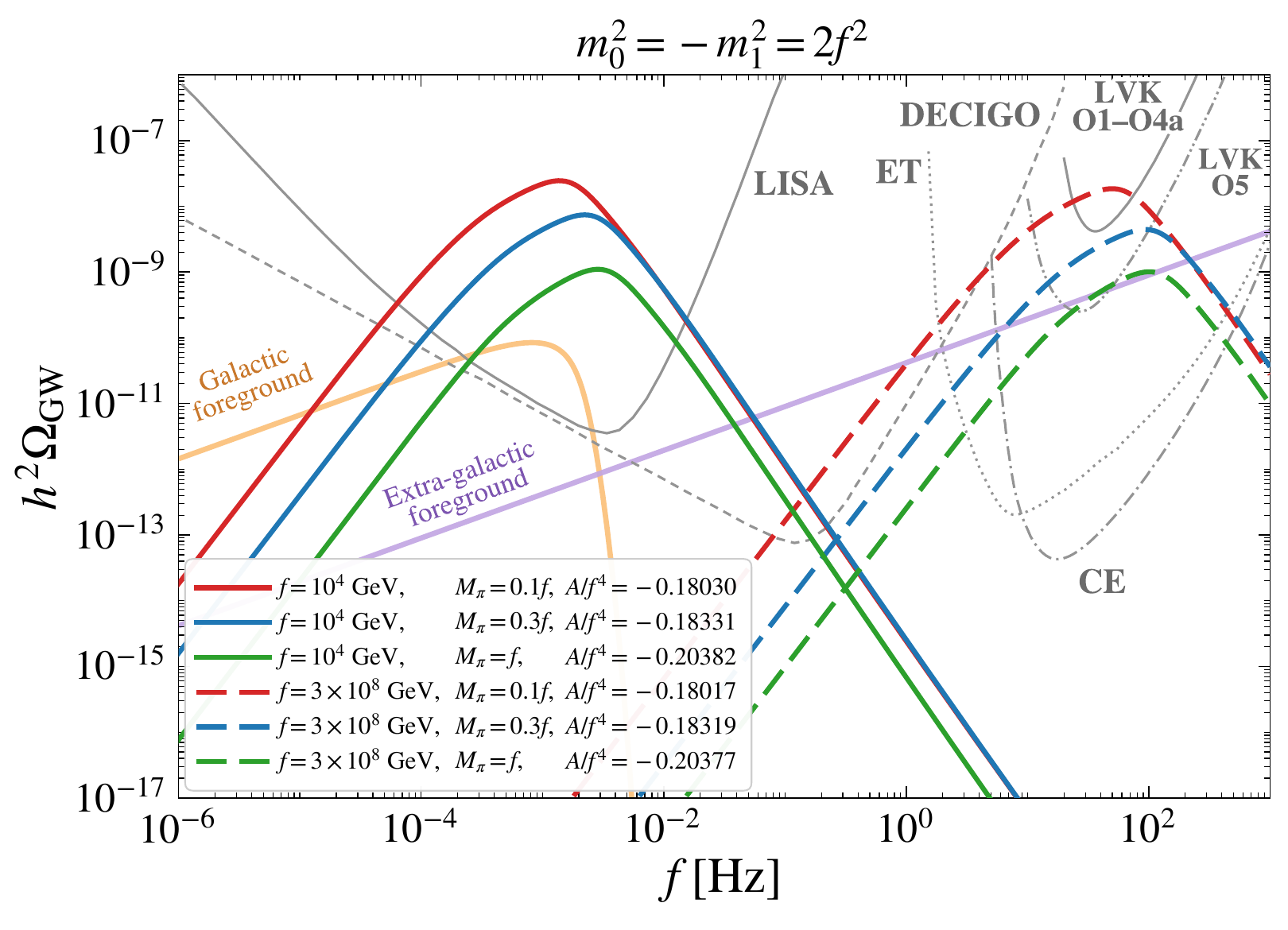}
    \caption{
    Prediction for the GW spectrum in each benchmark point. 
    The solid (dashed) colored lines indicate the signals for the case with $f = 10^{4}\,{\rm GeV}\,(3 \times 10^{8}\,{\rm GeV})$ and $\lambda_{\sigma} = 0.1$. 
    Gray lines are the sensitivity curves for LISA~\cite{LISA:2017pwj},  DECIGO~\cite{Kawamura:2020pcg}, Cosmic Explorer (CE)~\cite{Evans:2021gyd} and Einstein Telescope (ET)~\cite{ET:2025xjr}.
    For the LIGO-Virgo-KAGRA (LVK) collaboration~\cite{Virgo:2025aai}, the current constraint (O1--O4a) and the future expected sensitivity (O5) are shown.
    The light orange (purple) line represents the expected foreground from the inside (outside) of our galaxy~\cite{Caprini:2024hue}.  
    }
    \label{fig:GWprediction}
\end{figure}

Figure~\ref{fig:GWprediction} shows the predicted GW spectra for several benchmark points. We consider two representative values of the decay constant, $f=10^{4} \, {\rm GeV}$ and $f=3 \times 10^{8} \, {\rm GeV}$. The light orange (purple) line denotes the expected foreground from the inside (outside) of our galaxy~\cite{Caprini:2024hue}. For the case with $f=10^{4} \, {\rm GeV}$, the generated spectra have peak frequencies in the range relevant to future space-based GW detectors such as LISA~\cite{LISA:2017pwj} and DECIGO~\cite{Kawamura:2020pcg}. On the other hand, for $f=3 \times 10^{8} \, {\rm GeV}$, the corresponding GW signals may be explored by Cosmic Explorer~\cite{Evans:2021gyd} and Einstein Telescope~\cite{ET:2025xjr}.
Moreover, the benchmark point shown as the red dashed line is already constrained by the current bound from the data obtained by the LIGO-Virgo-KAGRA collaboration~\cite{Virgo:2025aai}. 
These results illustrate that our model can generate observable GW signals that are expected to be explored by current and future interferometers.

The peak frequency of the GW spectrum scales with $f$, while the peak amplitude and the required value of the parameter $A$ remain nearly unchanged. In particular, taking $f=\mathcal{O}(1) \,{\rm GeV}$ shifts the peak frequency into the nanohertz range accessible to pulsar timing array experiments~\cite{NANOGrav:2023gor,NANOGrav:2023hvm,EPTA:2023fyk,Reardon:2023gzh,Xu:2023wog}.

\section{Radiative Stability and Tuning}
\label{sec:tuning}

An important feature of our model is the radiative stability of the pNGB potential. As discussed in Sec.~\ref{sec:model}, collective symmetry breaking eliminates the quadratically divergent contributions from the $\sigma$ loops. The remaining logarithmic corrections to the cosine potential have a characteristic magnitude
$\delta A \sim {|m_0^2m_1^2|}/(16\pi^2)$.
The tree-level $A$ term, on the other hand, explicitly breaks the shift symmetry and generates additional radiative corrections. For $\Lambda\sim4\pi f$, the quadratically divergent correction from the $\phi$ loop is estimated to be of the same order as $A$ itself. The $\phi$ loops generate higher harmonics at logarithmic order, with coefficients of order $A^2/(16\pi^2f^4)$. Higher-loop corrections are expected to be suppressed by additional loop factors in the perturbative regime.

As indicated in Fig.~\ref{fig:alpha_beta}, realizing a small $\beta_{\rm eff}/H$ requires the $A$ parameter to be chosen close to a critical value, where its contribution to the curvature of the potential around $\phi=\pi f$ nearly cancels the $\sigma$-loop contribution. The magnitude $A/f^4\sim0.2$ is natural, since the tree-level coefficient is comparable to the radiative contribution. However, obtaining a sufficiently flat potential requires adjusting $A/f^4$ at the level of $\mathcal{O}(10^{-4})$. Compared with the typical size of radiative corrections, estimated to be $\mathcal{O}(10^{-2})$ for $m_0^2\sim -m_1^2\sim f^2$, this corresponds to a tuning of approximately $1\%$. The required tuning is therefore relatively mild.

\section{Conclusion}
\label{sec:conclusion}

In this paper, we have presented a simple and calculable model for realizing strong first-order phase transitions along a pNGB direction $\phi$. The interactions with an additional complex scalar field $\sigma$ exhibit collective symmetry breaking, protecting the pNGB potential against quadratically divergent corrections. A mass splitting between the two eigenstates of the $\sigma$ field generates a nonvanishing temperature-dependent potential when the heavier state becomes Boltzmann suppressed, allowing strong first-order phase transitions to occur.

Through a numerical analysis of the finite-temperature effective potential and bubble nucleation dynamics, we have identified parameter regions where phase transitions successfully complete with a large transition strength and a small inverse duration. The resulting GW signals can fall within the sensitivity ranges of future experiments. The model can also generate GWs in the nanohertz frequency range relevant to pulsar timing arrays.

\section*{Acknowledgment}

We thank Shota Nakagawa for fruitful discussions. 
YN is supported by Natural Science Foundation of Shanghai.

\bibliography{reference}
\end{document}